\documentstyle[emulateapj]{article}

\lefthead{Murayama et al.}
\righthead{NEW MID-INFRARED DIAGNOSTIC OF THE DUSTY TORUS MODEL FOR AGN} 

\begin{document}

\title{NEW MID-INFRARED DIAGNOSTIC OF THE DUSTY TORUS MODEL FOR SEYFERT NUCLEI}

\author{Takashi Murayama}
\affil{Astronomical Institute, Graduate School of Science, Tohoku University,
Aoba, Sendai 980-8578, Japan;
murayama@astr.tohoku.ac.jp}
\author{Hideaki Mouri}
\affil{Meteorological Research Institute,
1-1 Nagamine, Tsukuba 305-0052, Japan;
hmouri@mri-jma.go.jp}
\author{Yoshiaki Taniguchi}
\affil{Astronomical Institute, Graduate School of Science, Tohoku University,
Aoba, Sendai 980-8578, Japan;
tani@astr.tohoku.ac.jp}

\begin{abstract}

We propose a new diagnostic of the ``dusty torus'' model for Seyfert nuclei.
Dust grains in the torus are heated by the nuclear continuum, and reradiate
mostly in the mid-infrared wavelengths. From the torus geometry, it is
predicted that the emission at $\lambda \le 10$ \micron{} has strong
dependence on the viewing angle. Since the dependence is predicted to be
insignificant at $\lambda \ge 10$ \micron, we study the flux ratio between
3.5 \micron{} ($L$ band) and 25 \micron{}; $R(L,25) = \log [(\nu_{\rm
3.5\,\mu m}\,S_{\nu_{\rm 3.5\,\mu m}})/(\nu_{\rm 25\,\mu m}\,S_{\nu_{\rm
25\,\mu m}})]$. In three different samples (optically selected, X-ray
selected, and infrared selected samples) of Seyfert galaxies,
the observed values of $R(L,25)$ between type 1 Seyferts (S1s) and
type 2 Seyferts (S2s) are found to be
clearly separated; $R(L,25) > -0.6$ for S1s while $R(L,25) < -0.6$ for S2s.
This implies universality of their torus properties. With this result and
the other observational characteristics, we investigate the most plausible
torus model among those presented in Pier \& Krolik (1992, 1993).

\end{abstract}



\section{INTRODUCTION}
Dusty tori around active galactic nuclei (AGNs) play an important role
in the classification of Seyfert galaxies.
(Antonucci \& Miller 1985; see also Antonucci 1993 for a review).
Seyfert galaxies observed from a face-on view of the torus are
recognized as type 1 Seyferts (S1s) while those observed from a edge-on
view are recognized as type 2 Seyferts (S2s). In this way, the dusty tori
act as a material anisotropically obscuring the emission
from their interior region.

Dusty tori themselves are also important emitting sources in AGNs.
Dust grains within the torus absorb high-energy photons from the central
engine, and re-emit them in the mid-infrared (MIR) regime.
Therefore, infrared radiation from the dusty torus emission is
useful in studying the physical properties of the tori in AGNs
(e.g., Dopita et al.\ 1998 and references therein).
Since the tori are
quite optically thick, the MIR spectrum is predicted to have strong
dependence on the viewing angle [Efstathiou \& Rowan-Robinson 1990; Pier \&
Krolik 1992, 1993 (hereafter PK92 and PK93, respectively); Granato \&
Danese 1994; Granato, Danese, \& Franceschini 1996, 1997].
When the torus is observed from a face-on view, its hot inner
surface is seen and the emission at $\lambda \le$ 10 \micron{} is enhanced.
When observed from a edge-on view, the emission at $\lambda \le$ 10
\micron{} is obscured and thus weakened. Heckman (1995) observed that the
averaged ratio of $N$-band (10 \micron) flux to nonthermal radio flux is
higher in S1s than in S2s (see also Giuricin, Mardirossian, \& Mezzerre
1995). Heckman, Chambers, \& Postman (1992) observed a similar enhancement
in radio-loud quasars (i.e., type 1) with respect to radio galaxies (i.e.,
type 2). PK93 observed that flux ratios of $L$ band (3.5 \micron) to $N$
band in S1s are higher than those in S2s. Fadda et al.\ (1998) observed
that the MIR spectrum is steeper (i.e., redder) in S2s than in S1s.
However, further details of the MIR emission from dusty tori are unknown.

This paper proposes the flux ratio of $L$ band to {\it IRAS} 25 \micron{}
band as a new MIR diagnostic for the dusty torus model (\S2). We compile the
observational data from the literature (\S3), compare the above ratios of
S1s with those of S2s (\S4), and discuss properties of the tori (\S5).

\section{NEW MIR DIAGNOSTIC}
As stated above, the torus emission is expected to be more anisotropic at
$\lambda$ $\le$ 10 \micron{} than at $\lambda$ $\ge$ 20 \micron{}
because the visibility of the inner wall of the torus 
is highly viewing angle dependent.
Therefore, it is of interest to compare S1s with S2s in a
flux ratio between $\lambda \le 10$ \micron{} and $\lambda \ge 20$
\micron{}. Since the {\it IRAS} photometric data are available for most of
the nearby Seyfert galaxies (Moshir et al.\ 1992), we adopt the flux ratio
between $L$ band and {\it IRAS} 25 \micron{} band,
\[
R(L,25) = \log [(\nu_{\rm 3.5\,\mu m}\,S_{\nu_{\rm 3.5\,\mu m}})/ 
(\nu_{\rm 25\,\mu m}\,S_{\nu_{\rm 25\,\mu m}})]. 
\]
The basic concept of our new MIR diagnostic is schematically shown in Figure
1. Since the viewing angle dependence is more significant at 3.5 \micron{},
S2s are expected to have lower values of $R(L, 25)$ than S1s. 

\begin{figure*}
\epsscale{1.8}
\plotone{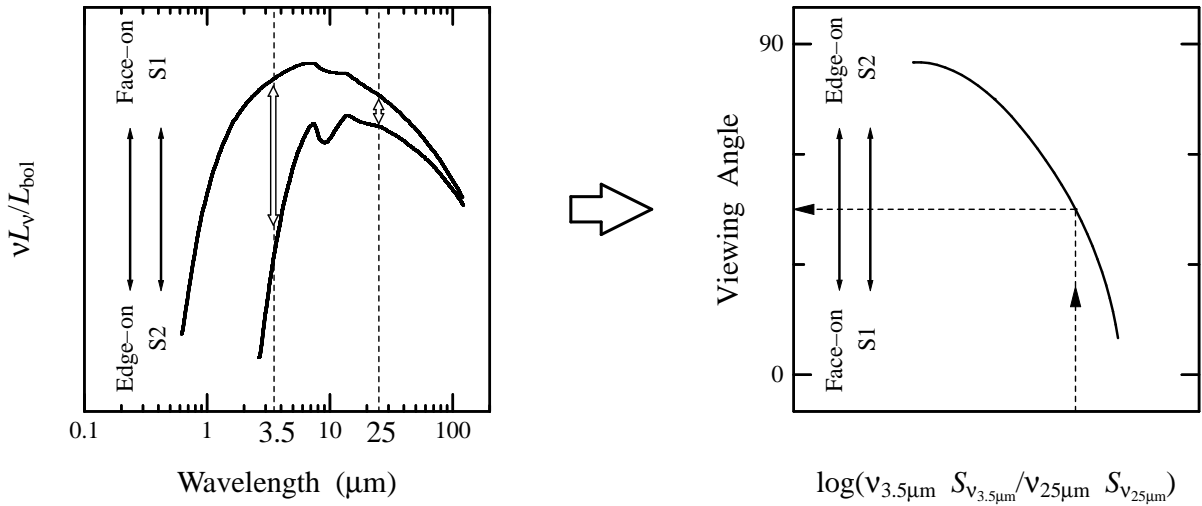}
\caption{Basic concept of our MIR diagnostic. The left panel shows
typical spectra of the torus emission for S1s (upper) and for S2s (lower).
The right panel shows how the 3.5 \micron{} to 25 \micron{} flux ratio
yields the viewing angle toward the torus.
\label{fig1}}
\end{figure*}

Here we note that PK 93 used the flux ratio between $L$ and $N$ bands,
$R(L, N)$, to compare S1s with S2s. However, the $N$-band flux is affected
by a silicate line at 9.7 \micron. When the torus is observed from a
edge-on view, the silicate line is seen as an absorption line, and thus
both the fluxes in $L$ and $N$ bands are weakened. The difference between
S1s and S2s in $R(L, N)$ is thereby expected not to be as prominent as that
in $R(L,25)$.

\section{DATA SAMPLE}
To perform a statistical analysis with the MIR diagnostic defined in the
previous section, we have compiled photometric data in $L$, $N$, and {\it
IRAS} 25 \micron{} bands from the literature (e.g., Ward et al.\ 1987;
Roche et al. 1991; Moshir et al.\ 1992; PK93). Since radiation form AGNs is
anisotropic in most of the energy bands, it is difficult to construct a
statistically complete sample. We instead adopt three samples chosen by
different selection criteria. The first sample consists of the CfA Seyfert
galaxies\footnote{%
The preliminary analysis based on the CfA Seyfert galaxies was reported in
Murayama, Mouri, \& Taniguchi (1997).}
(Huchra \& Burg 1992),
which provide a well defined collection of objects limited by the $B$
magnitude of their host galaxies. The second sample is the one limited by
the hard X-ray flux from 2 to 10 keV (Ward et al.\ 1987). Since hard X-rays
arise from the central engine itself and are not affected seriously by dust
grains, this sample is expected to be fair at least for S1s. The third
sample is taken from Roche et al.\ (1991).
This sample is not complete but composed of $N$-band bright objects.
For each object in this sample, Roche et al.\ (1991) observed an
emission feature at 11.3 \micron, which allows us to examine the
presence or absence of any circumnuclear star formation activities.
Our CfA, Ward, and Roche samples contain 18 S1s
and 6 S2s, 20 S1s and 4 S2s, and 11 S1s and 11 S2s, respectively. Some
objects are included in more than one sample. In total, there are 31 S1s
and 14 S2s. Their basic data are summarized in Table 1. 

\begin{deluxetable}{lcccccccccc}
\tablefontsize{\small}
\tablecaption{%
Infrared Properties of the Sample Seyfert Galaxies
}

\tablehead{
   \colhead{Name\tablenotemark{a}} & \colhead{$L$} & \colhead{$N$} &
   \colhead{$S_{\nu_{\rm 12\,\mu m}}$}  &
   \colhead{$S_{\nu_{\rm 25\,\mu m}}$}  &
   \colhead{$S_{\nu_{\rm 60\,\mu m}}$}  &
   \colhead{$S_{\nu_{\rm 100\,\mu m}}$} &
   \colhead{$C\!P$\tablenotemark{b}}    &
   \colhead{$R(L,25)$\tablenotemark{c}} &
   \colhead{Sample\tablenotemark{d}} &
   \colhead{Reject\tablenotemark{e}} \nl
 & \colhead{(Jy)} & \colhead{(Jy)} &
   \colhead{(Jy)} & \colhead{(Jy)} &
   \colhead{(Jy)} & \colhead{(Jy)} & & & &
}
\startdata
\multicolumn{11}{c}{Type 1 Seyfert} \nl
\hline
2237+07      & 0.016\phn      & 0.082      & \phn0.1400 & \phn0.3920 &
               \phn0.900      & \phn\phn1.270 &
               0.81           & $-0.54$    &
               C\phm{ W R}    &            \nl
3A 0557-383  & 0.113\phn      & 0.347      & \phn0.5289 & \phn0.6846 &
               \phn\phn0.3223 & \phn$<0.5589$ &
               0.78           & \phs0.07   &
               \phm{C }W R    &            \nl
3C 120       & 0.090\phn      & 0.220      & \phn0.2860 & \phn0.6350 &
               \phn1.283      & \phn\phn2.786  &
               1.00           & \phs0.00   &
               \phm{C }W R    &            \nl
Akn 120      & 0.0740         & 0.139      & \phn0.3191 & \phn0.4099 &
               \phn0.643      & \phn\phn1.084  &
               0.52           & \phs0.11   &
               \phm{C }W\phm{ R} &         \nl
ESO 141-G55  & 0.0640         & 0.184      & \phn0.2420 & \phn0.3522 &
               \phn\phn0.5741 & $<1.481$   &
               0.92           & \phs0.11   &
               \phm{C }W\phm{ R} &         \nl
I Zw 1       & 0.110\phn      & 0.390      & \phn0.5118 & 1.211      &
               \phn2.243      & \phn\phn2.643 &
               1              & $-0.19$    &
               C\phm{ W }R    &            \nl
IC 4329A     & 0.227\phn      & 0.760      & 1.082      & 2.213      &
               \phn2.030      & \phn\phn1.661  &
               0.90           & $-0.14$    &
               \phm{C }W R    &            \nl
MCG -6-30-15 & 0.0877         & 0.286      & \phn0.3803 & \phn0.8088 &
               \phn1.087      & \phn\phn1.096  &
               0.97           & $-0.12$    &
               \phm{C }W\phm{ R} &         \nl
MCG 8-11-11  & 0.075\phn      & 0.296      & \phn0.6394 & 1.948      &
               \phn3.005      & \phn\phn4.235  &
               0.65           & $-0.57$    &
               \phm{C }W R    &            \nl
Mrk 79       & 0.0543         & 0.200      & \phn0.3062 & \phn0.7625 &
               \phn1.503      & \phn\phn2.363  &
               0.87           & $-0.30$    &
               \phm{C }W\phm{ R} &         \nl
Mrk 231      & 0.360\phn      & 1.420      & 1.872      & 8.662      &
               31.99\phn      & 30.29      &
               1              & $-0.53$    &
               C\phm{ W }R    &            \nl
Mrk 279*     & 0.032\phn      & 0.076      & \phn0.1990 & \phn0.2890 &
               \phn1.200      & \phn\phn1.970 &
               0.46           & $-0.10$    &
               C\phm{ W R}    & $C\!P<0.5$ \nl
Mrk 335      & 0.123\phn      & 0.210      & \phn0.3021 & \phn0.3777 &
               \phn\phn0.3433 & \phn$<0.5673$  &
               0.83           & \phs0.37   &
               C\phm{ W R}    &            \nl
Mrk 509      & 0.113\phn      & 0.220      & \phn0.3158 & \phn0.7018 &
               \phn1.364      & \phn\phn1.521 &
               0.91           & \phs0.06   &
               \phm{C }W R    &            \nl
Mrk 530      & 0.025\phn      & 0.077      & \phn0.1800 & \phn0.1910 &
               \phn\phn0.8560 & \phn\phn2.140 &
               0.50           & $-0.03$    &
               C\phm{ W R}    &            \nl
Mrk 590      & 0.0469         & 0.169      & \phn0.1917 & \phn0.2214 &
               \phn\phn0.4893 & \phn\phn1.457 &
               1              & \phs0.18   &
               C W\phm{ R}    &            \nl
Mrk 766      & 0.0559         & 0.288      & \phn0.3855 & 1.295      &
               \phn4.026      & \phn\phn4.658 &
               1              & $-0.51$    &
               C\phm{ W R}    &            \nl
Mrk 817      & 0.0619         & 0.357      & \phn0.3350 & 1.175      &
               \phn2.118      & \phn\phn2.268 &
               1              & $-0.42$    &
               C\phm{ W R}    &            \nl
Mrk 841      & 0.0407         & \phn0.1445 & \phn0.1924 & \phn0.4726 &
               \phn\phn0.4593 & \phn$<0.6176$  &
               1              & $-0.21$    &
               C W\phm{ R}    &            \nl
Mrk 1040     & 0.105\phn      & 0.294      & \phn0.6104 & 1.315      &
               \phn2.555      & \phn\phn4.551  &
               0.63           & $-0.25$    &
               \phm{C }W\phm{ R} &         \nl
NGC 3227     & 0.0783         & 0.263      & \phn0.6671 & 1.764      &
               \phn7.825      & 17.59      &
               0.54           & $-0.50$    &
               C W\phm{ R}    &            \nl
NGC 3516     & 0.117\phn      & 0.239      & \phn0.4258 & \phn0.8937 &
               \phn1.758      & \phn\phn2.259 &
               0.73           & $-0.03$    &
               C\phm{ W R}    &            \nl
NGC 3783     & 0.131\phn      & 0.400      & \phn0.8396 & 2.492      &
               \phn3.257      & \phn\phn4.899  &
               0.67           & $-0.43$    &
               \phm{C }W R    &            \nl
NGC 4051*    & 0.077\phn      & 0.297      & \phn0.8554 & 1.590      &
               \phn7.131      & 23.92      &
               0.44           & $-0.46$    &
               C W R          & $C\!P<0.5$ \nl
NGC 4151     & 0.344\phn      & 1.400      & 2.080      & 4.600      &
               \phn6.720      & \phn\phn8.600 &
               0.88           & $-0.27$    &
               C W R          &            \nl
NGC 4593     & 0.081\phn      & 0.182      & \phn0.3441 & \phn0.8089 &
               \phn3.052      & \phn\phn5.947  &
               0.70           & $-0.15$    &
               \phm{C }W\phm{ R} &         \nl
NGC 5033*    & 0.0487         & 0.031      & \phn0.9452 & 1.148      &
               13.80\phn      & 43.85      &
               0.04           & $-0.52$    &
               C\phm{ W R}    & $C\!P<0.5$ \nl
NGC 5273     & 0.016\phn      & 0.134      & \phn0.1340 & \phn0.2420 &
               \phn\phn0.9910 & \phn\phn2.030 &
               1              & $-0.33$    &
               C\phm{ W R}    &            \nl
NGC 5548     & 0.0986         & 0.210      & \phn0.4006 & \phn0.7690 &
               \phn1.073      & \phn\phn1.614 &
               0.67           & $-0.04$    &
               C W\phm{ R}    &            \nl
NGC 7213     & 0.115\phn      & 0.261      & \phn0.6063 & \phn0.7421 &
               \phn2.666      & \phn\phn8.177  &
               0.51           & \phs0.04   &
               \phm{C }W\phm{ R} &         \nl
NGC 7469*    & 0.1594         & 0.600      & 1.348      & 5.789      &
               25.87\phn      & 34.90      &
               0.69           & $-0.71$    &
               C W R          & PAH        \nl
\hline
\multicolumn{11}{c}{Type 2 Seyfert} \nl
\hline
Circinus*    & 0.701\phn      & 6.00\phn   & 19.58\phn\phn & 71.29\phn\phn &
               248.7\phn\phn\phn & 315.9\phn\phn &
               0.45           & $-1.16$    &
               \phm{C W }R    & $C\!P<0.5$  \nl
Mrk 266      & 0.007\phn      & 0.306      & \phn0.2307 & \phn0.9765 &
               \phn7.432      & 11.07      &
               1              & $-1.29$&
               C\phm{ W R}    &             \nl
Mrk 348      & 0.039\phn      & 0.300      & \phn0.3080 & \phn0.8347 &
               \phn1.290      & \phn\phn1.549 &
               1              & $-0.48$    &
               \phm{C W }R    &             \nl
Mrk 533*     & 0.046\phn      & 0.217      & \phn0.6724 & 1.896      &
               \phn5.588      & \phn\phn8.146 &
               0.44           & $-0.76$    &
               C\phm{ W }R    & $C\!P<0.5$ \nl
NGC 1068     & 1.920\phn      & 18.0\phn\phn\phn  & 39.7\phn\phn\phn &
               85.04\phn\phn  & 176.2\phn\phn\phn & 224.0\phn\phn    &
               0.59           & $-0.79$    &
               C\phm{ W }R    &            \nl
NGC 1275     & 0.078\phn      & 0.674      & 1.069      & 3.539      &
               \phn7.146      & \phn\phn6.981 &
               0.91           & $-0.81$    &
               \phm{C W }R    &            \nl
NGC 2110*    & 0.047\phn      & 0.198      & \phn0.3488 & \phn0.8397 &
               \phn4.129      & \phn\phn5.676 &
               0.75           & $-0.40$    &
               \phm{C W }R    & NLXG       \nl
NGC 2992*    & 0.057\phn      & 0.249      & 0.594      & 1.422      &
               \phn6.941      & 14.44      &
               0.56           & $-0.55$    &
               \phm{C }W R    & PAH, NLXG \nl
NGC 3079*    & 0.073\phn      & 0.091      & 1.523      & 2.272      &
               44.5\phn\phn   & 89.2\phn   &
               0.07           & $-0.64$    &
               C\phm{ W R}    & $C\!P<0.5$ \nl
NGC 4388     & 0.074\phn      & 0.404      & \phn0.9964 & 3.463      &
               10.24\phn      & 18.10      &
               0.59           & $-0.82$    &
               C\phm{ W }R    &            \nl
NGC 5506*    & 0.313\phn      & 0.643      & 1.282      & 3.638      &
               \phn8.409      & \phn\phn8.886 &
               0.69           & $-0.22$    &
               \phm{C }W R    & NLXG       \nl
NGC 5929*    & 0.0073         & \phn0.0186 & 0.360      & 1.570      &
               \phn9.450      & 12.00      &
               0.08           & $-1.48$    &
               C\phm{ W R}    & $C\!P<0.5$ \nl
NGC 7172*    & 0.132\phn      & 0.141      & \phn0.4374 & \phn0.7612 &
               \phn5.712      & 12.29      &
               0.40           & \phs0.09   &
               \phm{C }W R    & $C\!P<0.5$, NLXG \nl
NGC 7582*    & 0.201\phn      & 0.877      & 1.620      & 6.436      &
               49.10\phn      & 72.92      &
               0.82           & $-0.66$    &
               \phm{C }W R    & PAH, NLXG \nl
\enddata
\tablenotetext{a}{Galaxies shown with asterisks are excluded from the analysis
                  because of possible contamination to the infrared emission}
\tablenotetext{b}{Compactness parameter defined in text}
\tablenotetext{c}{%
$R(L,25) = \log [(\nu_{\rm 3.5\,\mu m}\,S_{\nu_{\rm 3.5\,\mu m}})/
      (\nu_{\rm 25\,\mu m}\,S_{\nu_{\rm 25\,\mu m}})]$
}
\tablenotetext{d}{C: the CfA sample; W: the Ward sample;
                  R: the Roche sample}
\tablenotetext{e}{Reasons for excluding from the analysis}
\end{deluxetable}

Besides the dusty torus, several sources in Seyfert galaxies contribute to
the observed MIR fluxes (see below). We exclude galaxies where the
contamination with such sources appears to be significant. These galaxies
are indicated in Table 1. The resultant final samples consist of 27 S1s and
5 S2s.

The {\it IRAS} 25 \micron{} measurements were made with an aperture which
is large enough to cover the entire galaxy ($0\farcm{}75 \times
4\farcm{}6$; Neugebauer et al.\ 1984). There could be contamination with
the disk of the host galaxy. To find galaxies where the disk emission
dominates over the torus emission, we use the compactness parameter $C\!P$
at 10 \micron{} (Devereux 1987), 
$
C\!P = f_{\rm cc} \times S_{\nu_N}/S_{\nu_{\rm 12\,\mu m}}. 
$
Here $S_{\nu_N}$ is the $N$ band flux, $S_{\nu_{\rm 12\,\mu m}}$ is the
{\it IRAS} 12 \micron{} flux, and $f_{\rm cc}$ is the color correction
factor,
$
f_{\rm cc} = 0.12\,S_{\nu_{\rm 12\,\mu m}}/S_{\nu_{\rm 25\,\mu m}} + 1.04. 
$
This compactness parameter gives an estimate on the ratio of the small-beam
flux to the entire flux at 10 \micron. The $C\!P$ value of each galaxy is
given in Table 1. Some objects exhibit $C\!P > 1$. This is due to
uncertainties in the measurement or time variation of the nuclear flux. In
such cases, we give $C\!P=1$. The MIR fluxes of galaxies with $C\!P < 0.5$
are likely to be dominated by the disk emission. These galaxies are not
used in our following analysis.

On the other hand, the $L$- and $N$-band data given in Table 1 were
obtained with small apertures ($\phi \simeq$ 5\arcsec--10\arcsec). In these
data, the contamination with the host galaxy is unlikely to be important.
From $K$-band images of Seyfert galaxies, Kotilainen et al.\ (1992)
estimated the average light contribution from the host galaxy as 32 \%.
Zitelli et al.\ (1993) found that $L$-band images of Seyfert galaxies are
more centrally concentrated than the $K$-band ones. Hence the contribution
from the host galaxy to the $L$-band flux is less than $\sim 30$ \%.

Seyfert galaxies often exhibit circumnuclear starburst activities, which
could affect the MIR emission (see Keto et al.\ 1992 for the case of NGC
7469). Such objects are excluded from our analysis. As a signature of the
starburst activity, we use emission features in the 8--13 \micron{} regime
(Roche et al.\ 1991). They are due to transient heating of polycyclic
aromatic hydrocarbon molecules (PAHs) by UV photons from OB stars. Since
PAHs are destroyed by X-rays, PAH features are absent in genuine AGNs
(see Voit 1992 and references therein).

We also exclude narrow-line X-ray galaxies (NLXGs), i.e., S2s with strong
hard X-ray emission (Shuder 1980; V\'eron et al.\ 1980; Ward et al.\ 1987).
The central engine of
these galaxies is believed to be hidden not by a dusty tori but by the disk
of the host galaxy. Most of NLXGs are actually edge-on galaxies (see
Ulvestad \& Wilson 1984 and Keel 1980 for the cases of NGC 2992 and NGC
5506). Furthermore, Glass et al.\ (1981) reported that $R(L,N)$ values of
NLXGs are similar to those of S1s rather than those of S2s.

\section{RESULTS}
Figure 2 shows frequency distributions of $R(L,25)$ of S1s and S2s
separately for the CfA sample ($a$), the Ward sample ($b$), the Roche
sample ($c$), and the total sample ($d$). The galaxies excluded in the
previous section are shown by white bars. Only the galaxies shown by black
bars are used in the following analysis.

\begin{figure*}
\epsscale{0.8}
\plotone{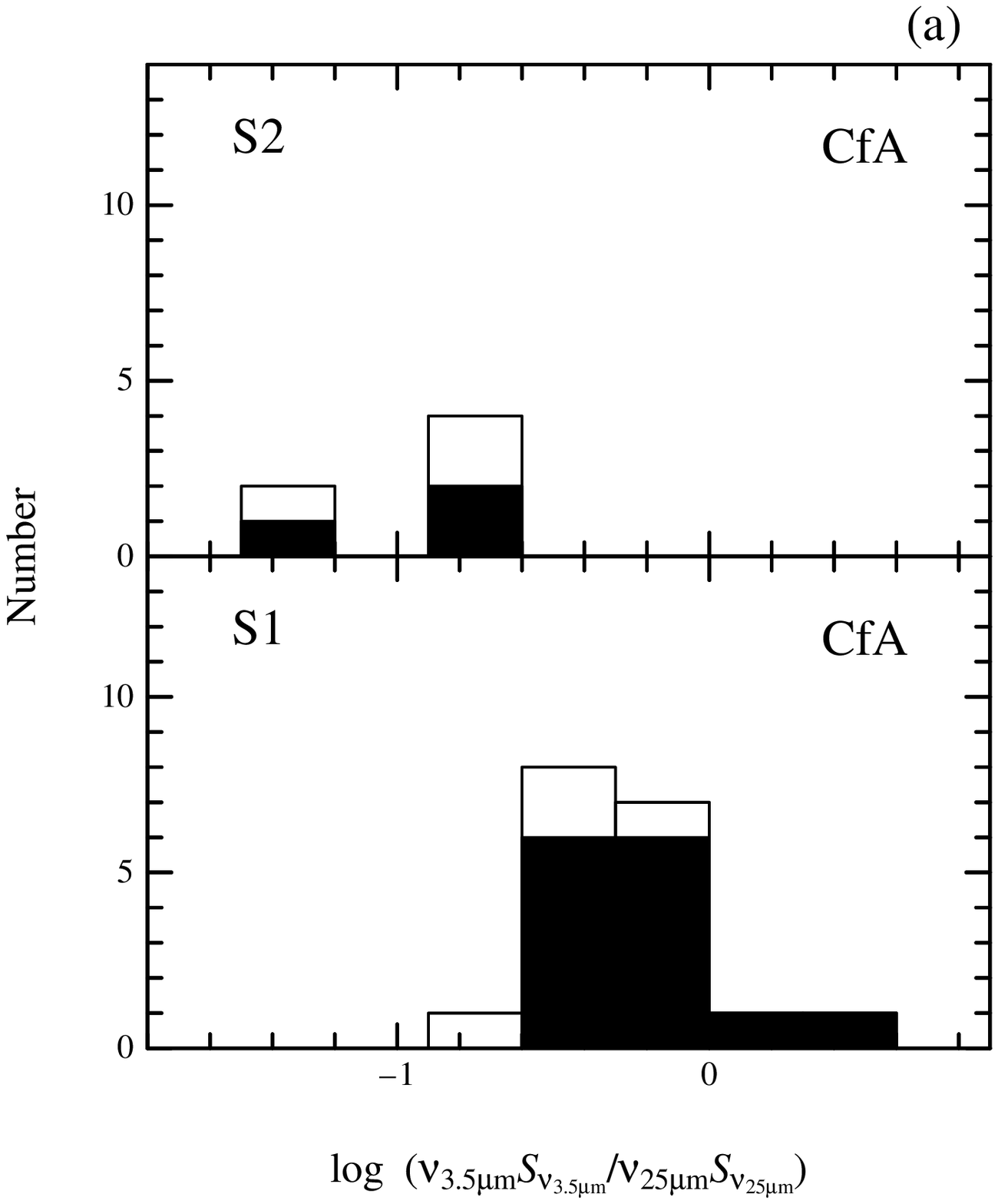}
\plotone{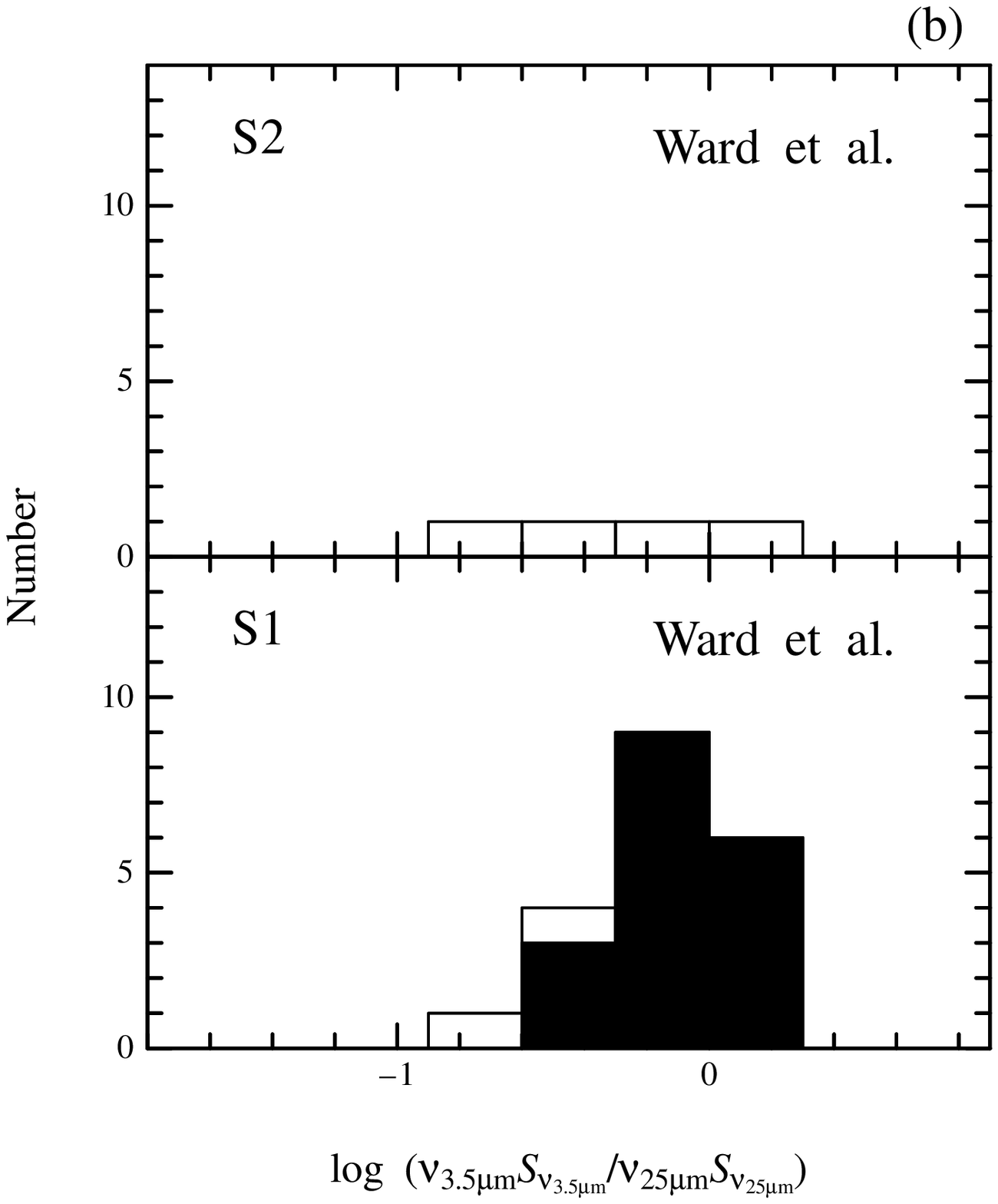}\\
\plotone{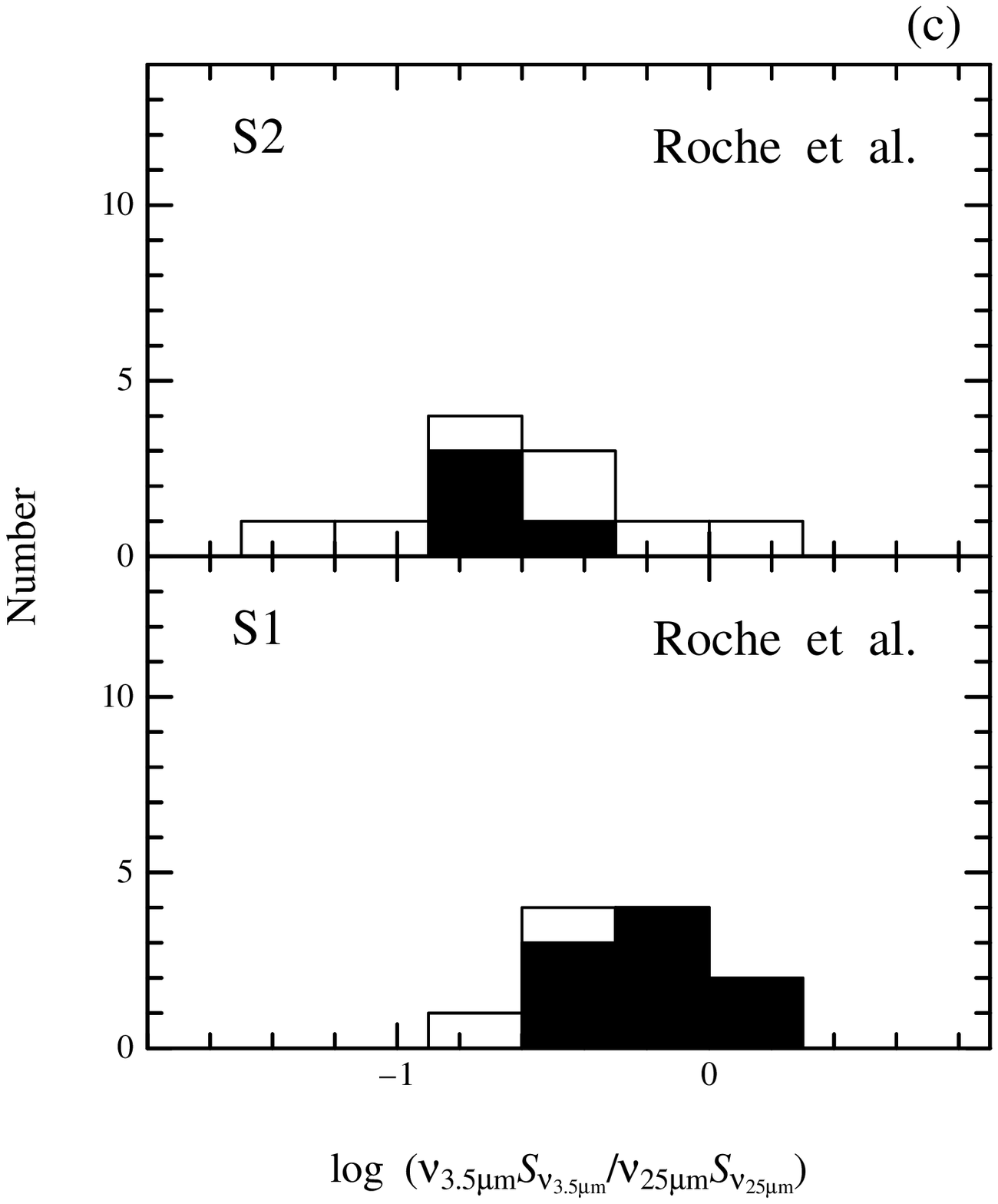}
\plotone{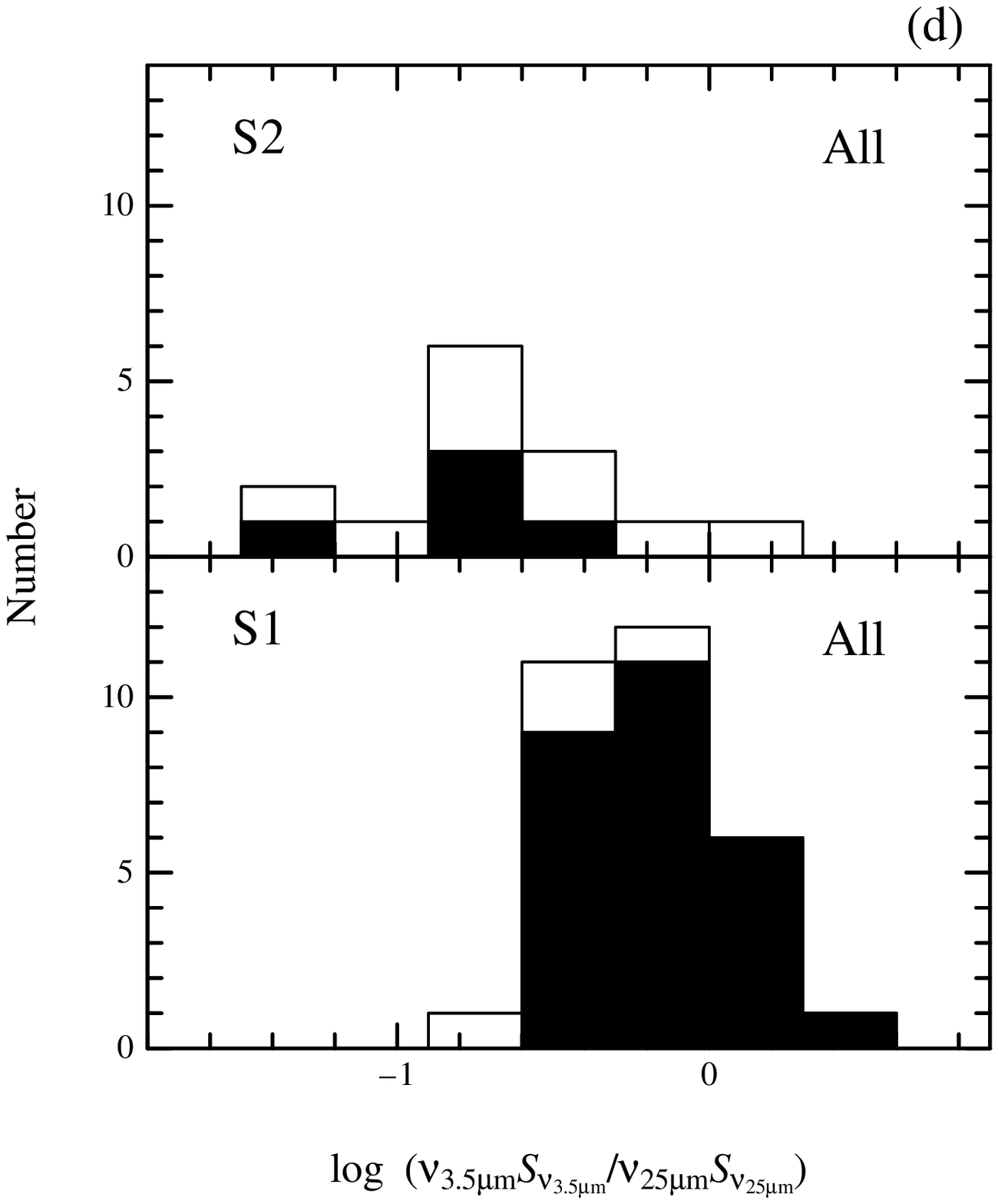}
\caption{Histogram of $R(L,25)$ for the CfA Seyferts
($a$), the sample of Ward et al.\ (1987) ($b$), the sample of Roche et al.\
(1991) ($c$), and the total sample ($d$). Galaxies shown by white bars are
likely to suffer from contamination and are not used in our analysis (see
text and Table 1).
\label{fig2}}
\end{figure*}

All of the S1s have $R(L,25) > -0.6$ while most of the S2s have $R(L,25) <
-0.6$. The S2 which lies exceptionally at $R(L,25) > -0.6$ is Mrk 348. This
galaxy exhibits no silicate absorption feature at 9.7 \micron{} (Roche et
al.\ 1991). Since the silicate absorption is a common property of S2s, Mrk
348 is considered to be in a face-on view like usual S1s. The absence of
the broad-line region in this galaxy could result from, e.g., obscuration
of the central region by a small cloud. There is no significant
difference in the distributions of S1s and S2s among the three
samples in Figures 2$a$--$c$,
thus our samples is  probably free of large orientation bias.
If we apply the Kolmogrov-Smirnov (KS) test, the probability that
the observed distributions of S1s and S2s originate in the same
underlying population
turns out to be 0.275 \%. When galaxies shown by white bars are included,
the distribution of S2s is different among the three samples. This
difference is likely to come from the different sampling criteria.

Figure 3 compares the $R(L,25)$ ratio with the nuclear absolute $B$
magnitude [$M_B$ (nucleus)] for S1s.
The $M_B$ (nucleus) values are taken from Kotilainen, Ward, \&
Williger (1993) and Granato et al.\ (1993), and are used as a measure
of the luminosity of the central engine.
If the observed $R(L,25)$ ratio depends on the
intrinsic luminosity of the central engine rather than the Seyfert type,
there would be a certain relationship between $R(L,25)$ and $M_B$
(nucleus). Since no clear correlation is seen in Figure 3, we conclude that
the difference in the intrinsic nuclear luminosity does not affect the
observed value of $R(L,25)$ in S1s. This conclusion is applicable to S2s
because S1s and S2s are likely to have the same torus properties.

\begin{figure*}
\epsscale{1.2}
\plotone{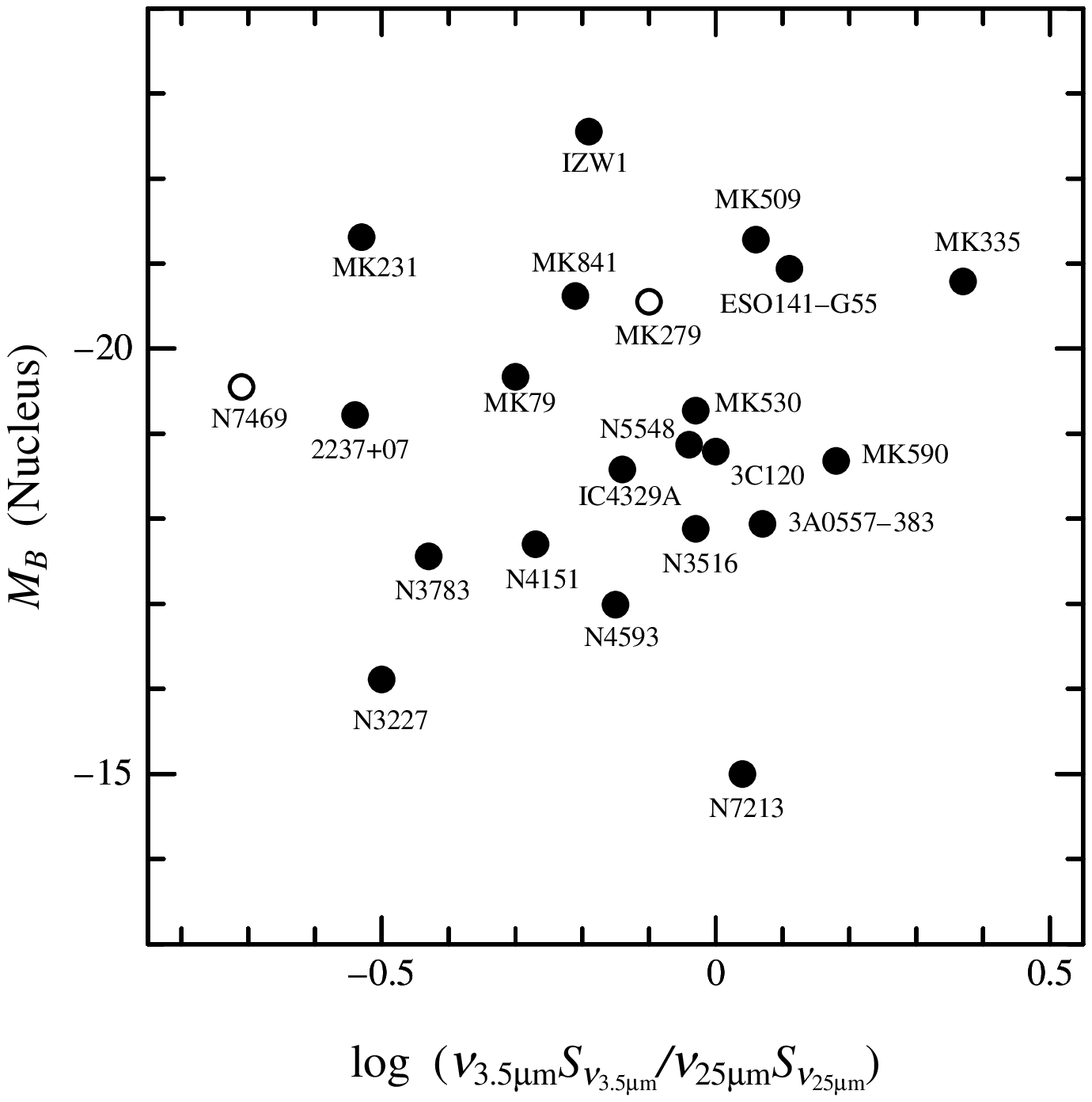}
\caption{%
Diagram of $R(L,25)$ vs.\ absolute $B$ magnitude of nuclei for S1s.
Filled circles represent galaxies used in our analysis.
Two open circles are Mrk 279 and NGC 7469, which are included
in the sample but not used in our analysis (see text and Table 1).
Nuclear $B$ magnitudes were taken from Kotilainen, Ward, \& Williger
(1993) and Granato et al.\ (1993) which measured magnitudes of
unresolved nuclear components deconvolved from two-dimensional CCD
images of Seyfert galaxies.
The Hubble constant of 75 km s$^{-1}$ Mpc$^{-1}$ is assumed
to obtain the absolute magnitudes.
\label{fig3}}
\end{figure*}

Finally, we show the frequency distributions of $R(L,N)$ in Figure 4. The
KS probability that the underlying populations of S1s and S2s are the same
is 0.0404 \%. Although this value is smaller than that for the $R(L,25)$
ratio, the separation between S1s and S2s in $R(L,N)$ (Figure 4) is less
clear than that in $R(L,25)$ (Figure 2). This is because the $N$-band
emission is affected by the silicate feature at 9.7 \micron.

\begin{figure*}
\epsscale{1.2}
\plotone{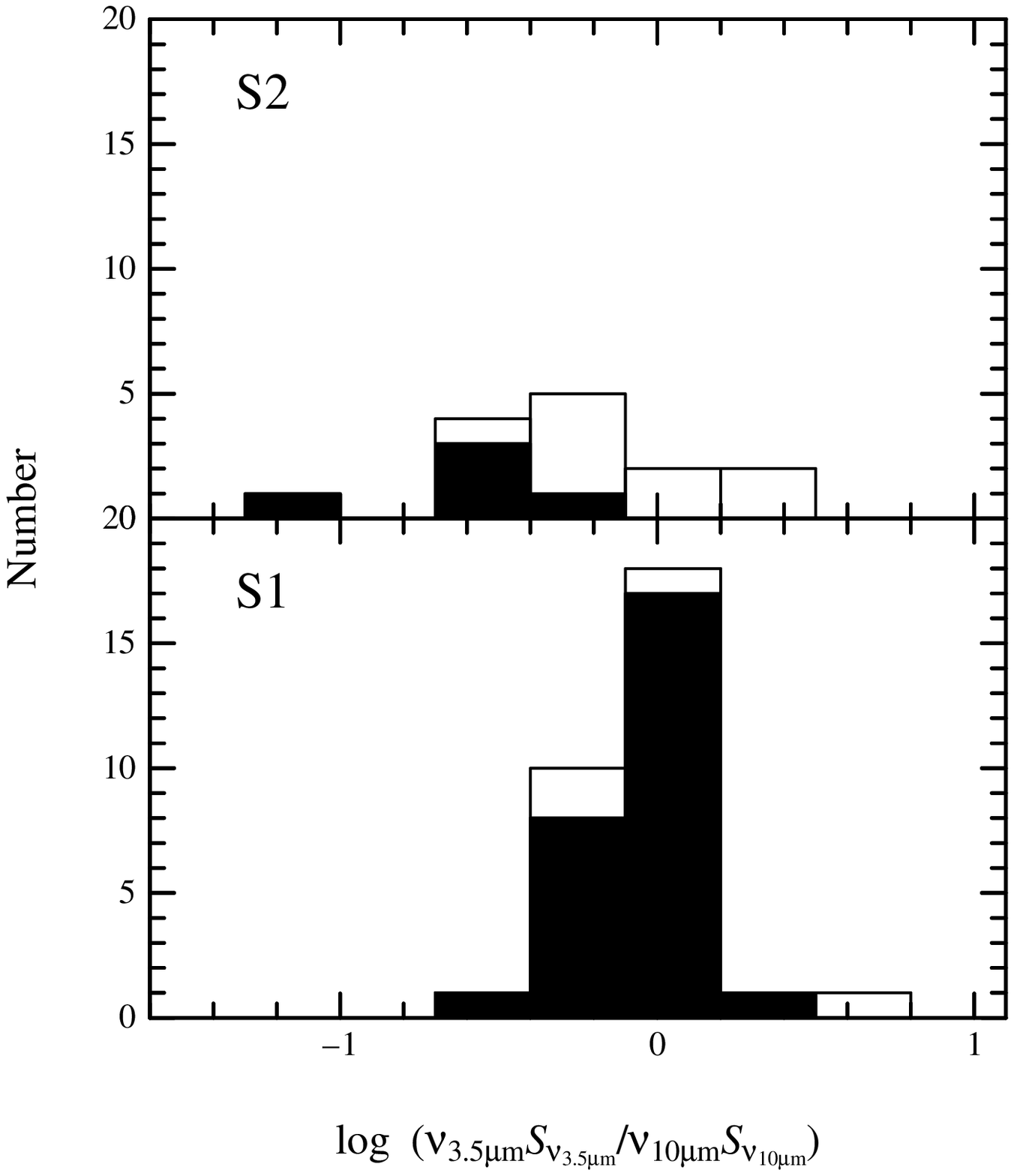}
\caption{Frequency distributions of $R(L,N)$ for the total sample.
\label{fig4}}
\end{figure*}

\section{DISCUSSION}
The $R(L,25)$ values for the S1s are clearly separated from
those for the S2s at the critical value of $-0.6$.
This limits the extent to which the dusty torus can vary among
Seyfert galaxies;
such variations would add ``noise'' and cause overlap between the two types.
Hereafter, we compare our results with theoretical torus models of
PK92 and PK93, and investigate the model which agrees best with
the observations. 

Figure 5 shows the geometrical configuration assumed in PK92 and PK93. The
torus surrounds cylindrically around the central engine and the broad-line
region. The semi-opening angle $\theta_{\rm open}$ is given by the inner
radius $a$ and the height $h$ of the torus; $\theta_{\rm open}=\tan^{-1}
(2a/h)$. The viewing angle $i$ is defined as an angle between the rotation
axis of the torus and the line of sight. The critical viewing angle $i_{\rm
cr}$ is defined such that the broad-line region is visible at $i < i_{\rm
cr}$. Since the actual torus should be clumpy and not have a sharp edge, we
expect $i_{\rm cr} \ge \theta_{\rm open}$.

\begin{figure*}
\epsscale{1.2}
\plotone{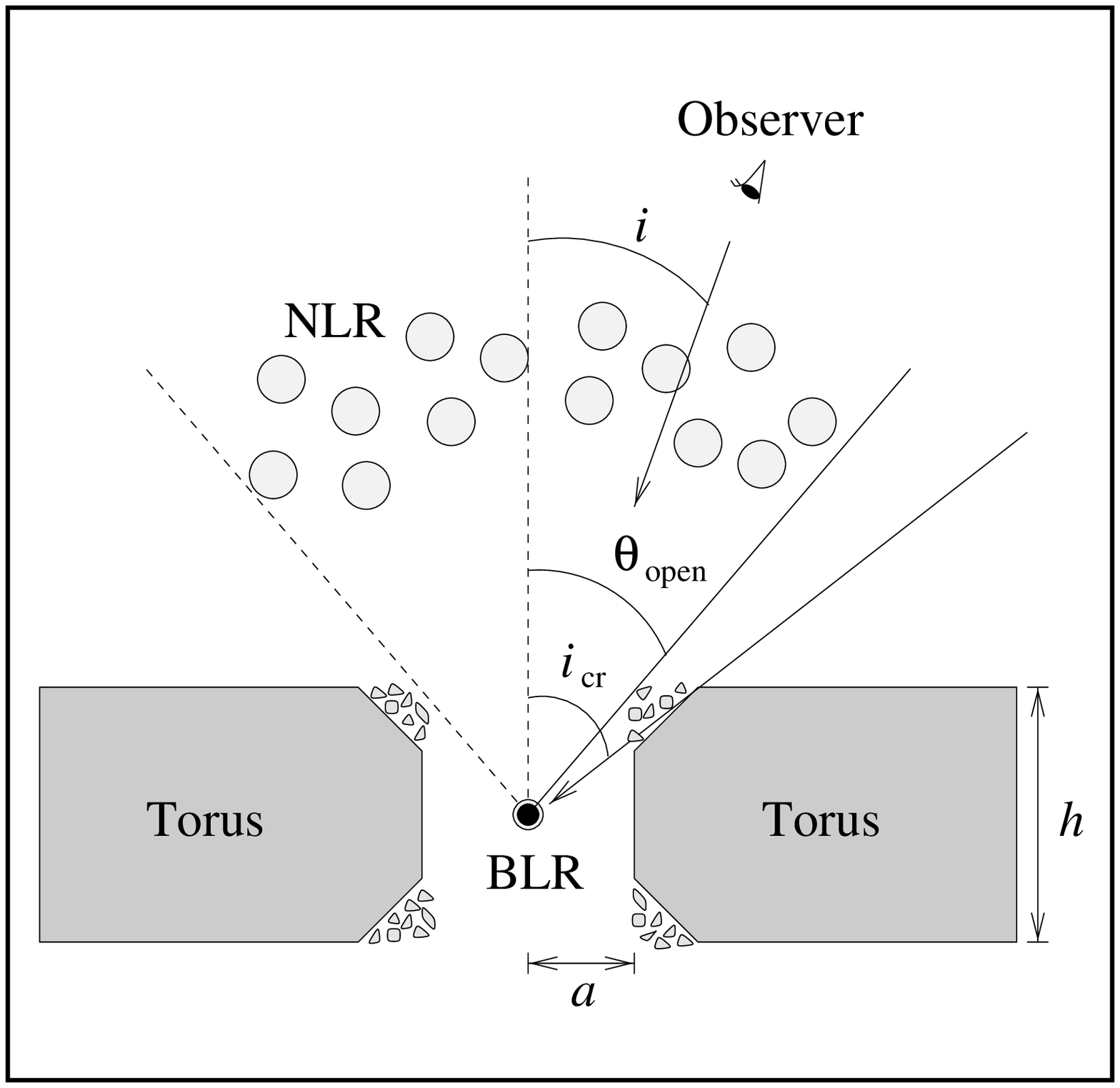}
\caption{Schematic illustration of the geometrical configuration of the
dusty torus model.
\label{fig5}}
\end{figure*}

The torus emission is parameterized by the three quantities
in the models of PK92 and PK93; 1) $T$: the
effective temperature of the inner wall of the torus, 2) $a/h$: the inner
aspect ratio, and 3) $\tau_{\rm r}$ and $\tau_{\rm z}$: the radial and
vertical Thomson optical depths.
In the upper panel of Figure 6, we show
the theoretical $R(L,25)$ values as a function of the viewing angle $i$ for
six dusty torus models of PK92 and PK93. In the lower panel, the observed
$R(L,25)$ values are shown separately for S1s and S2s. For each of the
models, the observed critical ratio, $R(L,25) = -0.6$, yields a critical
viewing angle. The results together with the model parameters are given in
Table 2.

\begin{deluxetable}{cccccccc}
\tablecaption{%
The Dusty Torus Models and the Derived Critical Viewing Angles
}
\tablehead{
  \colhead{Model} &
  \colhead{$T$\tablenotemark{a}} &
  \colhead{$a/h$\tablenotemark{b}} &
  \colhead{$\tau_{\rm r}$\tablenotemark{c}} &
  \colhead{$\tau_{\rm z}$\tablenotemark{d}} &
  \colhead{$\theta_{\rm torus}$\tablenotemark{e}} &
  \colhead{$i_{\rm cr}$\tablenotemark{f}} \nl
  & (K) &   &  &  & (\arcdeg) & (\arcdeg)
}
\startdata
1 &    1000 & 0.1 & 1   & 1   & 11 & 44 \nl
2 &    1000 & 0.3 & 0.1 & 0.1 & 31 & \nodata \nl
3 &    1000 & 0.3 & 1   & 0.1 & 31 & \nodata \nl
4 &    1000 & 0.3 & 1   & 1   & 31 & 87 \nl
5 & \phn800 & 0.3 & 1   & 1   & 31 & 82 \nl
6 & \phn500 & 0.3 & 1   & 1   & 31 & 46 \nl
\enddata
\tablenotetext{a}{Effective temperature of the torus}
\tablenotetext{b}{Inner aspect ratio of the torus}
\tablenotetext{c}{Radial Thomson optical depth of the torus}
\tablenotetext{d}{Vertical Thomson optical depth of the torus}
\tablenotetext{e}{Semi-opening angle derived from the inner aspect ratio}
\tablenotetext{f}{Expected viewing angle of the torus}
\end{deluxetable}

The derived critical viewing angle ranges from 46\arcdeg{} to 87\arcdeg{}
(Models 1, 4, 5, and 6). Since Models 2 and 3 do not give the critical
viewing angle, these models are not appropriate for dusty tori
in Seyfert galaxies.
To proceed further, we have to compare the results with
other observational properties of
Seyfert galaxies (see below).

Narrow-line regions of S2s often exhibit conical morphologies, which are
due to shadowing of the nuclear ionizing continuum by the torus. The
observed semi-opening angle of the cone $\theta_{\rm open} ({\rm NLR})$ is
thereby equal to the semi-opening angle of the torus $\theta_{\rm open}$.
Table 3 summarizes statistical results from observations of conical
narrow-line regions (Pogge 1989; Wilson \& Tsvetanov 1994; Schmitt \&
Kinney 1996). These results indicate $\theta_{\rm open} ({\rm NLR}) \simeq
30\arcdeg$. On the other hand, Model 1 has $\theta_{\rm open} = 11\arcdeg$.
Thus this model is not appropriate for dusty tori in Seyfert galaxies.


\begin{deluxetable}{lcc}
\tablecaption{%
Semi-Opening Angles  of Ionization Cones of Seyfert Nuclei
}
\tablehead{%
  \colhead{Reference} &
  \colhead{$N_{\rm Seyfert}$\tablenotemark{a}} &
  \colhead{$\theta_{\rm open} ({\rm NLR})$}
}
\startdata
Pogge 1989                & \phn4 & $26\arcdeg \pm    11\arcdeg$ \nl
Wilson and Tsvetanov 1994 &    11 & $32\arcdeg \pm \phn8\arcdeg$ \nl
Schmitt and Kinney 1996   &    12 & $29\arcdeg \pm \phn9\arcdeg$ \nl
\enddata
\tablenotetext{a}{Number of the observed Seyfert galaxies}
\end{deluxetable}

The critical viewing angle $i$ can be estimated from the number statistics
of S1s and S2s if we observe Seyfert nuclei from random orientations on
the statistical ground,
\[
{N({\rm S1}) \over {N({\rm S1}) + N({\rm S2})}} = 1 - \cos i_{\rm cr} \:
{\rm (stat)},
\]
where $N$(S1) and $N$(S2) are the observed numbers of S1s and S2s,
respectively (Miller \& Goodrich 1990).
Table 4 summarizes the results for three different surveys of
Seyfert galaxies (Osterbrock \& Shaw 1988; Salzer 1989; Huchra \& Burg
1992). The derived critical viewing angles ranges from 27\arcdeg{} to
46\arcdeg. Since Models 4 and 5 give too large critical viewing angles
($i_{\rm cr} > 80\arcdeg$), they are not appropriate for dusty tori in
Seyfert galaxies. Consequently, among the six models of PK92 and PK93,
Model 6 with $\theta_{\rm open} = 31\arcdeg$ and $i_{\rm cr} = 46 \arcdeg$
is the best torus model. 

\begin{deluxetable}{lcccc}
\tablecaption{%
Semi-Opening Angles Derived from the Statistics of Type 1 and
Type2 Seyfert Galaxies
}
\tablehead{%
  \colhead{Reference} &
  \colhead{$N_{\rm S1}$\tablenotemark{a}} &
  \colhead{$N_{\rm S2}$\tablenotemark{b}} &
  \colhead{$f_{1/2}$\tablenotemark{c}} &
  \colhead{$i_{\rm cr} ({\rm stat})$}
}
\startdata
Osterbrock and Shaw 1988 & \phn6 & \phn9 & 0.125    & 27\arcdeg \nl
Salzer 1989              & \phn9 & \phn7 & 0.20\phn & 34\arcdeg \nl
Huchra and Burg 1992     &    25 &    23 & 0.435    & 46\arcdeg \nl
\enddata
\tablenotetext{a}{Number of type 1 Seyfert galaxies}
\tablenotetext{b}{Number of type 2 Seyfert galaxies}
\tablenotetext{c}{Number ratio between type 1 and
                  type 2 Seyfert galaxies corrected for
                  the completeness of the survey}
\end{deluxetable}

The $R(L, 25)$ values of the S2s lie between $-1.48$ and $-0.6$, which
correspond to the viewing angles between $86\arcdeg$ and $46\arcdeg$. On
the other hand,  $R(L, 25)$ values of the S1s lie between $-0.6$ and
$0.37$. This range is not explained by Model 6. The locus of Model 6 in
Figure 6 is drawn down only to $i=41\arcdeg$. Since the $R(L, 25)$ value at
the smaller viewing angle is expected to be nearly constant, it would be
impossible to reproduce the $R(L, 25)$ ratio as high as $0.37$.
One possibility that explains this higher ratio may be
the 3 \micron{} bump often seen in type 1 AGNs.
This bump may be attributed to thermal emission
from hot dust grains with $T \simeq 1300$ K (e.g., PK93).
Because the models of PK92 and PK93 assumed that the host
dust component is an additional source to the torus,
$R(L, 25)$ is underpredicted at small inclination angles.
Although it is controversial whether this host dust
component is a separate component from the torus or the
inner surface of the torus, those hot dust grains
lie close to the central engine in either case.
Since their emission is important only when the central region is
clearly visible,
the 3 \micron{} bump is negligible in S2s.
Therefore the critical  $R(L,25)$ and $i_{\rm cr}$ 
are not affected by the treatment of the 3 \micron{} bump.

\begin{figure*}
\epsscale{1.2}
\plotone{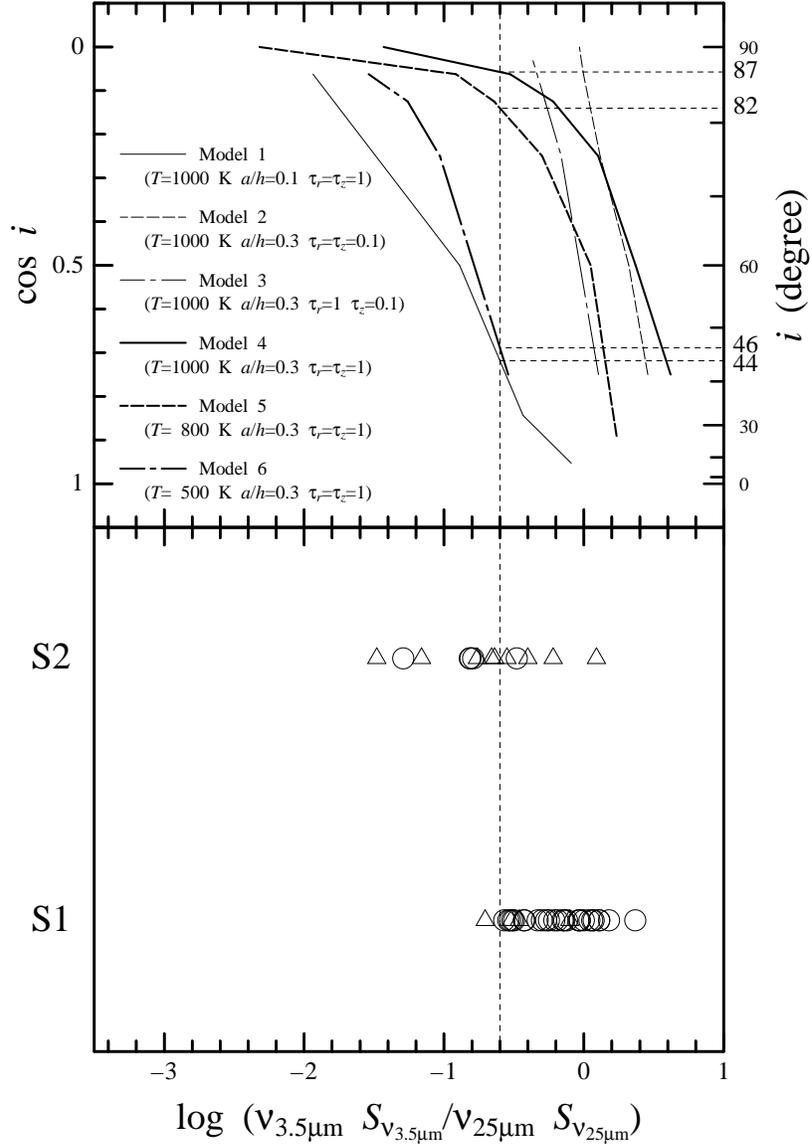}
\caption{Upper panel: relationships between $R(L,25)$ and the viewing
angle for six dusty torus models given in Table 4. Lower panel:
distributions of the observed $R(L,25)$ values. Galaxies shown by open
triangles are likely to suffer from contamination and are not used in our
analysis (see text and Table 1).
\label{fig6}}

\end{figure*}

We have examined only the small and sparse sets of the model parameters
presented by PK92 and PK93. Further analyses with the larger and denser
parameter sets are required to understand the torus properties in more
detail. Nevertheless, our most important result, which has been obtained
firstly with the new MIR diagnostic, is the clear separation in the $R(L,25)$
ratio between S1s and S2s. This strongly suggests that the torus properties
do not vary among Seyfert galaxies.
The effective temperature of the inner wall may be universal
as a result of that the inner wall is formed
by balancing the rate of dust destruction with the rate at which
the torus clouds drift inward (Krolik \& Begelman 1988; PK92).
We suspect that there are also
certain mechanisms confining the vertical structure of the torus and
shaping the uniform semi-opening angle.

\acknowledgments
We thank K.\ Iwasawa and R.\ Antonucci for helpful comments and suggestions
that improved the paper. TM is a JSPS Research Fellow. This work was
financially supported in part by Grant-in-Aids for the Scientific Research
(Nos.\ 07044054, 10044052, and 10304013) of the Japanese Ministry of
Education, Science, Sports, and Culture.

\clearpage

\end{document}